\documentclass[twocolumn,showpacs,preprintnumbers,amsmath,amssymb,superscriptaddress]{revtex4}
\pdfoutput=1

\usepackage[pdftex]{graphicx}
\usepackage[pdftex,colorlinks=true,citecolor=blue,urlcolor=blue,linkcolor=black]{hyperref} 

\usepackage{graphicx}
\usepackage{dcolumn}
\usepackage{bm}
\usepackage{datetime}
\usepackage{ulem}
\usepackage{float}

\begin{document}

\title{Matter wave Fourier optics with a strongly interacting two-dimensional Fermi gas } 

\author{P. A. Murthy}
\email{murthy@physi.uni-heidelberg.de}
\affiliation{Physikalisches Institut, Ruprecht-Karls-Universit\"at Heidelberg, 69120 Heidelberg, Germany}

\author{D. Kedar}
\affiliation{Physikalisches Institut, Ruprecht-Karls-Universit\"at Heidelberg, 69120 Heidelberg, Germany}

\author{T. Lompe}
\altaffiliation[Current address: ]{MIT-Harvard Center for Ultracold Atoms, 
Massachusetts Institute of Technology, Cambridge, Massachusetts
02139, USA }
\affiliation{Physikalisches Institut, Ruprecht-Karls-Universit\"at Heidelberg, 69120 Heidelberg, Germany}

\author{M. Neidig}
\affiliation{Physikalisches Institut, Ruprecht-Karls-Universit\"at Heidelberg, 69120 Heidelberg, Germany}

\author{M. G. Ries}
\affiliation{Physikalisches Institut, Ruprecht-Karls-Universit\"at Heidelberg, 69120 Heidelberg, Germany}

\author{A. N. Wenz}
\affiliation{Physikalisches Institut, Ruprecht-Karls-Universit\"at Heidelberg, 69120 Heidelberg, Germany}

\author{G. Z\"urn}
\affiliation{Physikalisches Institut, Ruprecht-Karls-Universit\"at Heidelberg, 69120 Heidelberg, Germany}

\author{S. Jochim}
\affiliation{Physikalisches Institut, Ruprecht-Karls-Universit\"at Heidelberg, 69120 Heidelberg, Germany}

\date{\today \, at \currenttime} 

\begin{abstract}
We demonstrate and characterize an experimental technique to directly image the momentum distribution of a strongly interacting two-dimensional quantum gas with high momentum resolution. We apply the principles of Fourier optics to investigate three main operations on the expanding gas: focusing, collimation and magnification. We focus the gas in the radial plane using a harmonic confining potential and thus gain access to the momentum distribution. We pulse a different harmonic potential to stop the rapid axial expansion which allows us to image the momentum distribution with high resolution. Additionally, we propose a method to magnify the mapped momentum distribution to access interesting momentum scales. All these techniques can be applied to a wide range of experiments and in particular to study many-body phases of quantum gases. 

\end{abstract}

\maketitle

\section{Introduction}
In a many-body quantum system, the distributions of particles in position and momentum space contain complementary information on the state of the system. 
In many cases, the change in the characteristic properties of the system at a phase transition is more pronounced in momentum space. Prominent examples for this are Bose-Einstein Condensation \cite{Anderson1995,Davis1995}, the Berezinskii-Kosterlitz-Thouless (BKT) transition \cite{Hadzibabic2006} and the Superfluid to Mott-Insulator transition \cite{Bloch2002}. Therefore, it is desirable to not only observe the in-situ density distribution of the system but also to access the momentum distribution.

In ultracold quantum gas experiments, this can often be achieved by performing conventional time-of-flight (TOF) imaging, i.e.  switching off all trapping potentials and letting the gas expand for a certain time $t$ before imaging \cite{ketterle1999}. 
In this process, the particles expand according to their initial momentum and thus the momentum distribution can be obtained from the density distribution after the time-of-flight. There are however several limitations to this technique. First, a direct mapping from momentum to position coordinates is only possible in the so-called far field limit which is reached for $t \rightarrow \infty$. Only then the influence of the initial distribution of the sample vanishes. 
However, in an experiment the maximum feasible TOF is usually limited by a decreasing signal-to-noise ratio and distortions due to residual potentials.
A second major challenge is that, for a strongly interacting system, interatomic collisions during the expansion can cause a significant redistribution of momentum. For such a non-ballistic expansion, the obtained spatial distribution does not reflect the initial momentum distribution of the sample. 
Therefore, to access the true momentum distribution of a strongly interacting quantum gas, it is crucial to develop methods that overcome the limitations of this technique. So far the best candidate for weakly interacting systems has been the Bose-gas focusing technique \cite{Shvarchuck2002, VanAmerongen2008,Jacqmin2012,VanEs2010,Tung2010} which brings the far field limit to finite TOF. In \cite{Tung2010} this is achieved by letting the cloud expand in a weak harmonic potential. 
However all these methods crucially rely on a ballistic expansion  of the sample, which is challenging to achieve for strongly interacting systems.

In this paper, we build upon this work to develop techniques for using Fourier optics of matter waves to perform three different tasks: focusing, collimation and magnification. In the following sections, we first provide an explanation of the working principle behind the technique and then describe how we use the operations of focusing and collimation to measure the momentum distribution of a strongly interacting 2D Fermi gas. We then propose a magnification scheme which allows to measure the momentum distribution of the system with high resolution.

\section{Working principle} 

The ballistic expansion of an ultracold gas can be understood in close analogy with the far field limit in Fourier optics. Using the Fraunhofer diffraction model, the field distribution at a large distance from the source is the Fourier transform of the initial complex field. In the case that a parabolic lens is placed in the optical path, it imprints a quadratic phase shift on the complex field and thus brings the far field distribution to its focal plane \cite{Saleh1991}. Here, our objective is to implement an equivalent matter wave lens which brings the far field to experimentally accessible time-scales. 
Note that the far field in this case is defined not in terms of distance from the source, but rather by the time elapsed after releasing the particle from the trap.

\begin{figure} [ht!]
\includegraphics [width=7.cm] {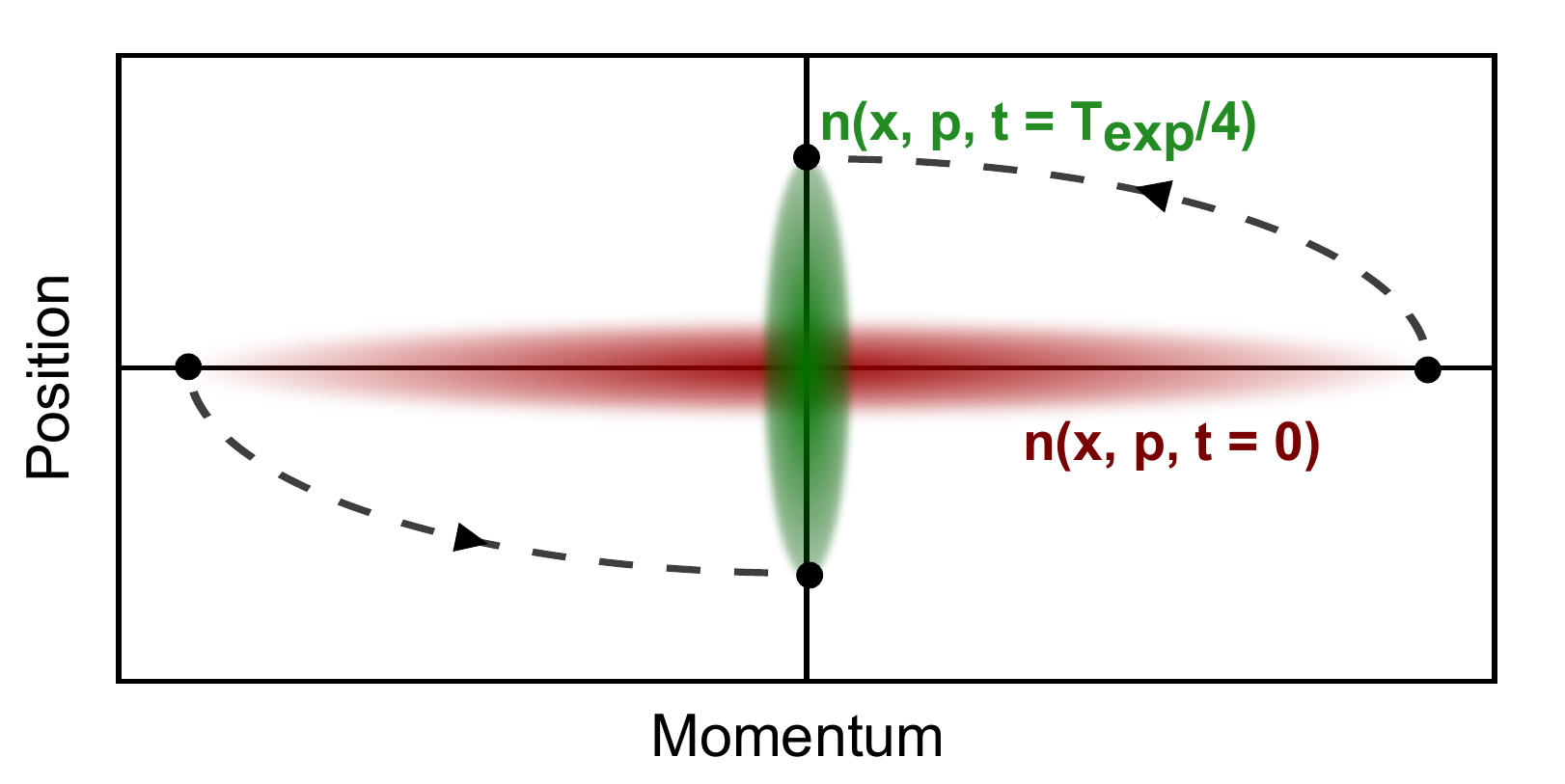}
\caption{Mapping between momentum and position space in phase space representation. The dashed lines illustrate the elliptic phase space trajectories of non-interacting particles in a harmonic potential. The phase space distributions $n(x,p)$ at $t=0$ and $t=T_{\text{exp}}/4$ are shown in red and green.} 
\label{fig:phasespace}
\end{figure}

Creating a matter wave lens can be accomplished by letting the gas evolve in a harmonic potential $V_{\text{exp}}(x) = \frac{1}{2}m\omega_{\text{exp}}^2x^2$ instead of a conventional TOF expansion. Here, $m$ is the mass of particles and $\omega_{\text{exp}}$ is the harmonic oscillator frequency. Without loss of generality, we will only consider a one-dimensional system. This is justified since in a harmonic potential and in the absence of interactions the equations of motion are separable. In a classical picture, each particle starts with some initial momentum and undergoes simple harmonic motion in the potential applied during the expansion. After a quarter of the time period of $V_{\text{exp}}(x)$, i.e. $t = \frac{T_{\text{exp}}}{4} = \frac{1}{4}\frac{2\pi}{\omega_{\text{exp}}} $, the position of each particle is directly proportional to its initial momentum. As illustrated in Fig.\ref{fig:phasespace}, each particle travels along an elliptical trajectory in phase-space and therefore the entire phase-space distribution undergoes a $\pi/2$ rotation after $T_{\text{exp}}/4$.

This classical model can be extended to a description using quantum fields. Let us consider the quantum field operators $\hat{\Psi}(x) $ and $ \hat{\Psi}^{\dagger}(x) $ \cite{Negele1998}, which respectively annihilate and create a particle in a particular quantum state at a point $x$ \footnote{Depending on the quantum statistics of the system, the field operators obey certain commutation relations. The statistics does not affect the outcome of the technique.}. The density distribution of particles in position and momentum space are then given by
\begin{equation}
n(x) =  \langle \hat{\Psi}^\dagger (x)\hat{\Psi}(x) \rangle, \quad
n(p) =  \langle \tilde{\Psi}^\dagger (p)\tilde{\Psi}(p) \rangle. 
\label{densitydist}
\end{equation}
In the Heisenberg picture, the evolution of the field operator with a Hamiltonian $\hat{H}$ is governed by 
\begin{equation}
i\hbar \partial_t \hat{\Psi}(x, t) = [\hat{\Psi}(x, t),\hat{H}].
\label{HeisenbergEq1}
\end{equation}
If we assume a ballistic expansion in a harmonic potential, i.e. without interactions during the expansion, Eq.\ref{HeisenbergEq1} reduces to
\begin{align}
i \hbar \, \partial_t \hat{\Psi}(x, t) = \Big[-\frac{\hbar^2}{2m}\nabla^2 + \frac{1}{2}m\omega_{\text{exp}}^2x^2 \Big]\hat{\Psi}(x, t). 
\label{Schrodinger}
\end{align}
This time-evolution equation has the form of the Schr\"{o}dinger equation and is linear in the field. It is in close analogy to the paraxial wave equation in optics which describes the spatial propagation of electromagnetic fields. 
Solving the field equation, it can be shown that at $T_{\text{exp}}/4$, the spatial field operator reflects the initial field operator in momentum space. Therefore, by measuring the spatial density distribution at $t=T_{\text{exp}}/4$, one can infer the momentum distribution at $t=0$ according to
\begin{align}
n(x, t= T_{\text{exp}}/4)&= \langle \hat{\Psi}^\dagger (x, T_{\text{exp}}/4)\hat{\Psi}(x, T_{\text{exp}}/4) \rangle \nonumber \\
          &= \langle \tilde{\Psi}^\dagger ({p},0)\tilde{\Psi}({p},0) \rangle = n(p,t=0),
\label{mapping}
\end{align}
where $p = m\omega x$. A detailed proof can be found in appendix A. From this description, it is apparent that the harmonic potential brings the far field to a single ``focal plane'' which is realized at an expansion time $t=T_{\text{exp}}/4$. 

It is important to note that the mapping between position and momentum space (Eq.\ref{mapping}) in this method only works because of the quadratic structure of the Hamiltonian (Eq.\ref{Hamiltonian}). This means that interparticle interactions during the expansion would alter the final spatial distribution resulting in a distorted momentum distribution. Hence in an experiment, it is crucial to ensure that interparticle interactions play only a negligible role during the expansion of the gas. In the following section we describe the experimental realization of this technique using a 2D Fermi gas in the BEC-BCS crossover and describe how we overcome the issues arising from strong interactions.

\section{Experimental realization} 

\label{sect_focusing}
For our experiments we use an ultracold Fermi gas of $^6$Li atoms in the lowest two Zeeman sublevels ($\vert 1 \rangle $=$\vert F$=$\frac{1}{2},m_F$=$-\frac{1}{2}\rangle$ and $\vert 2 \rangle$=$\vert \frac{1}{2}, \frac{1}{2}\rangle$ \cite{Gehm2003}). We bring this gas into the 2D regime by loading it into a hybrid trap consisting of an optical standing-wave in z-direction and a weak magnetic confinement in radial direction. The trapping frequencies of this combined potential are $\omega_{\text{z}}$=$2\pi \times 5.5$\,kHz in axial and $\omega_{\text{x,y}}$=$2\pi \times 18$\,Hz in radial direction. 
To tune the strength of the interparticle interactions in the gas, we vary the 3D s-wave scattering length $a_{3D}$. This is achieved by applying magnetic offset fields close to a broad Feshbach resonance centered at $832\,$G \cite{Zuern2013}. We probe the system by measuring the atomic density distribution in the $x$-$y$ plane by performing resonant absorption imaging along the direction of strong confinement ($z$-axis). The details of the experimental setup and the preparation scheme can be found in \cite{Wenz2013,Ries2014}.


\subsection{Focusing}
\begin{figure}[ht!]
\includegraphics [width=8cm]{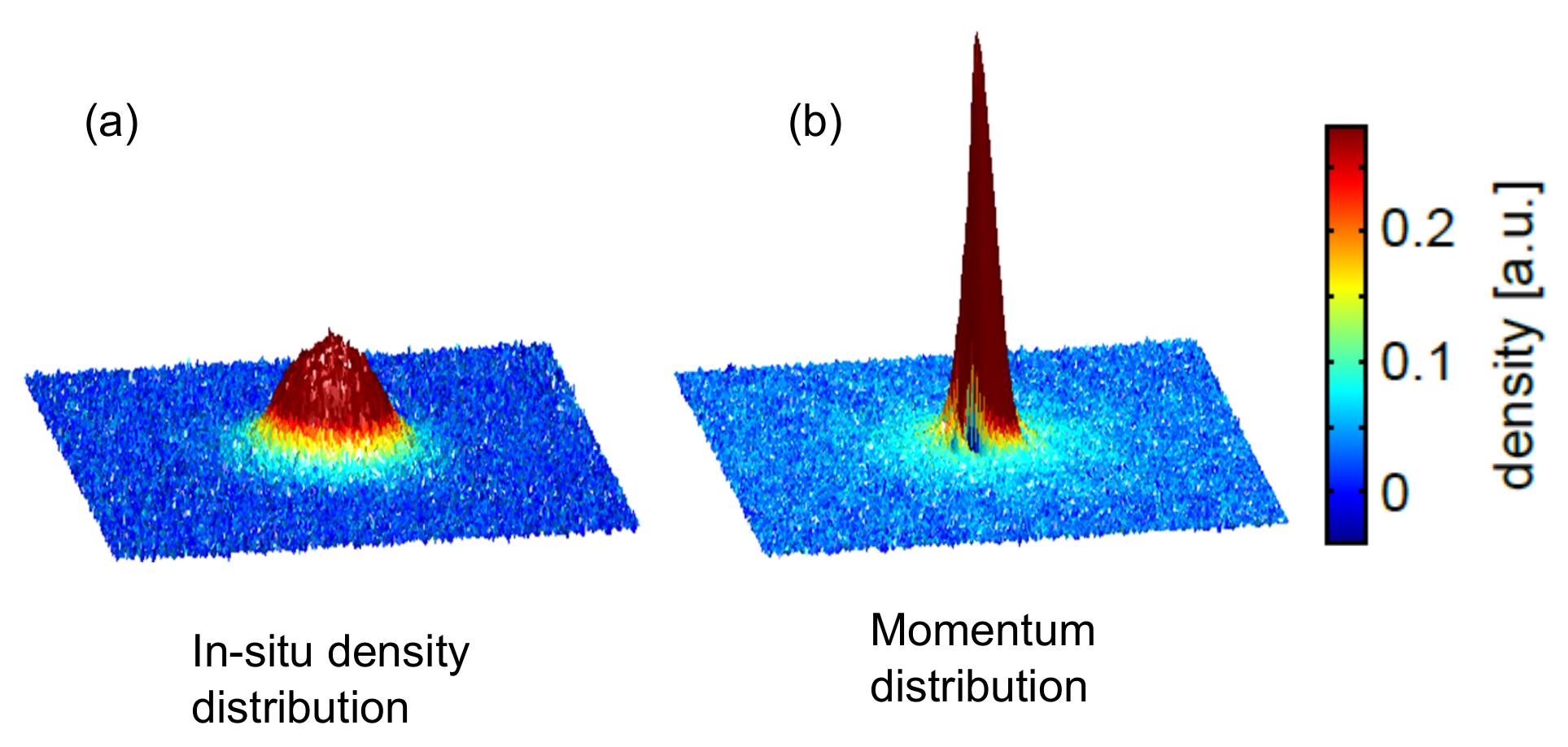}
\caption{Experimentally determined in-situ density distribution (a) and the momentum distribution (b) of a 2D gas at a magnetic field of $B = 692\,$G where $a_{3D}\simeq 1000\,a_{\text{Bohr}}$. The momentum distribution of the gas is obtained from the mapping to a density distribution using the focusing technique described in section \ref{sect_focusing}. At sufficiently low temperatures, we observe an enhanced occupation of low momentum states which is not apparent in the in-situ density distribution. A detailed investigation of this phenomenon will be reported elsewhere \cite{Ries2014}.}
\label{fig:focusing}
\end{figure}

In order to access the momentum distribution we realize the technique theoretically discussed above in a way similar to the one presented in \cite{Tung2010} for a 2D Bose gas: we switch off the optical trap and let the gas evolve in the weak magnetic potential, which has a harmonic trapping frequency $\omega_{\text{exp}} = 2\pi \times 10$\,Hz in radial direction \footnote{Magnetic potentials are well suited for this application since they are usually very smooth and have low anharmonicitiy, which is equivalent to a matter wave lens which has only minimal aberrations.}. As the strong confinement along the $z$-direction is switched off the sample rapidly expands in z-direction and quickly enters the ballistic regime. After a time-of-flight of $t = T_{\text{exp}}/4 = 25$\,ms the $x$-$y$ momentum distribution at $t=0$ has been mapped to a density distribution which we observe with absorption imaging. As an example, Fig.\ref{fig:focusing} shows images of the in-situ density distribution and the corresponding momentum distribution obtained with this focusing technique averaged over about $30$ experimental realizations. 

Using the rapid expansion along the z-direction to bring the sample into the ballistic regime works well for weak to intermediate interaction strengths. For example, at a magnetic field of 692\,G, we expect less than 0.1 scattering events per particle on average during the expansion (see appendix B). However, in the strongly interacting regime, scattering events during the expansion can significantly distort the mapping of the momentum distribution. We solve this problem by performing a magnetic field ramp with a duration of less than $150$\,$\mu$s to a field where the scattering length is smaller, just before release. This ramp converts pairs of atoms into deeply bound molecules, which allows us to measure the pair momentum distribution of the sample \cite{Regal2004,Zwierlein2004,Zwierlein2005}. 
The combination of the interaction quench by the rapid ramp and the quick expansion in z-direction leads to a ballistic expansion of the sample even in the strongly interacting regime.
We can therefore use this technique to measure the momentum distribution of our 2D Fermi system across the whole BEC-BCS crossover \cite{Levinsen2014}.

\subsection{Collimation}
While the rapid expansion in the axial direction ensures a quick reduction of the density and thus ballistic expansion, it also introduces a limitation for imaging the momentum distribution: 
During the T/4 expansion in the magnetic potential, the axial size of the cloud grows strongly and can exceed the depth of focus of the imaging setup. This limits the optical resolution, and therefore the momentum resolution. After the fast initial expansion, we thus need to limit the axial size of the gas to have sufficient resolution after long TOF.

\begin{figure}[ht!]
\includegraphics [width=8cm]{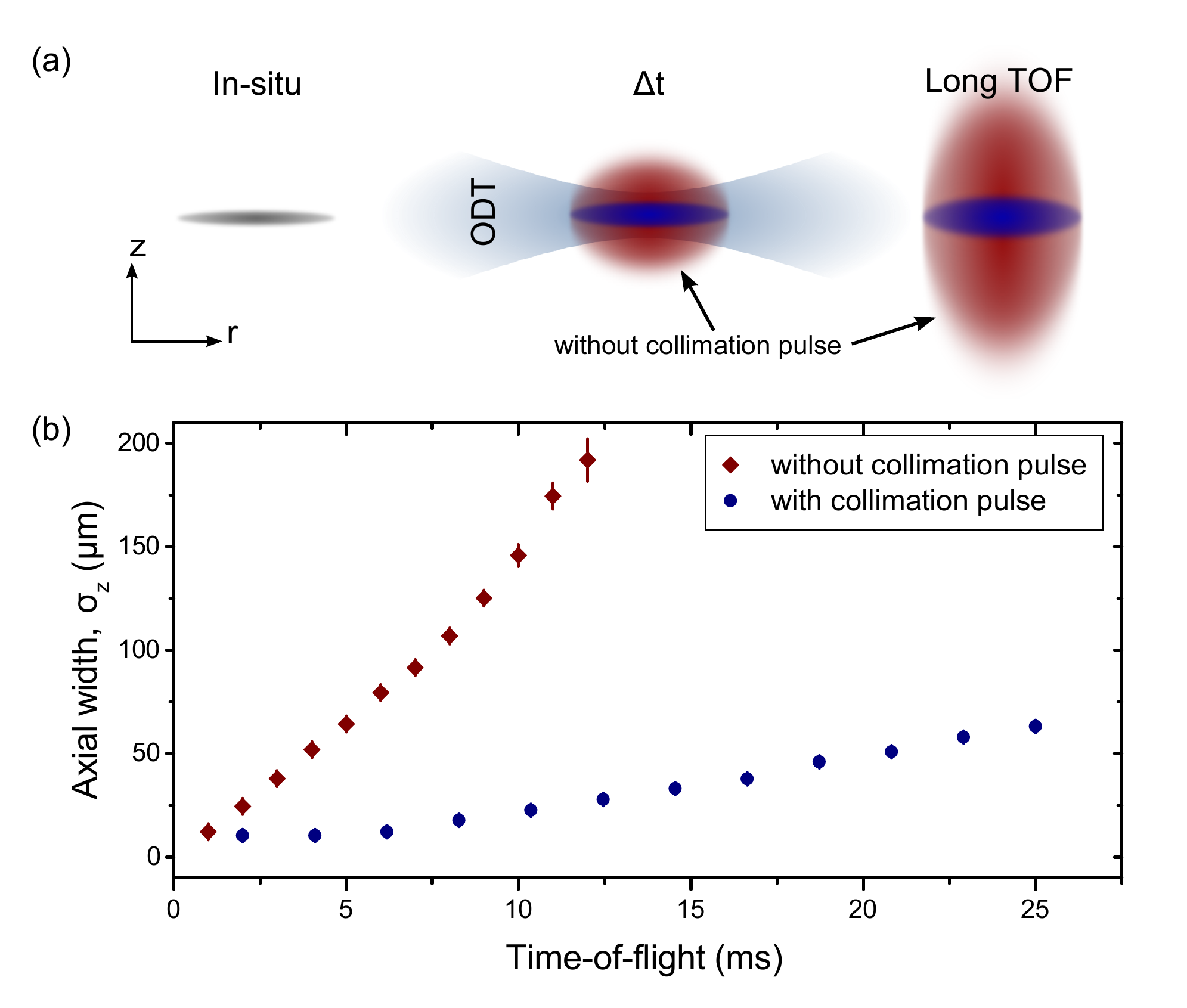}
\caption{(a) Schematic of the axial expansion with and without the collimation pulse. The initial cloud (gray) expands strongly in axial direction when the collimation pulse is not applied (red). With the collimation pulse however, the axial expansion is significantly slowed down (blue). (b) Measurement of axial width as a function of TOF with (blue) and without (red) collimation pulse. In this case the duration of the collimation pulse is $\Delta t_{\text{col}}=0.5\,$ms.}
\label{collimation}
\end{figure}

We achieve this by using the matter wave Fourier optics concept again, this time to collimate the expanding cloud in the axial direction. After the 2D trapping potential is turned off, we switch on an approximately harmonic optical dipole trap potential (ODT). The ODT is switched off when the particles have reached the classical turning point in the potential, i.e. $\Delta t_{\text{col}} = {T_{\text{ODT}}}/{4} = {\pi}/{2 \omega_{\text{ODT}}}$. This means that once the collimation pulse is turned off, the axial expansion should be stopped and only the radial motion persists (Fig.\ref{collimation}a). This method is similar to the Delta-Kick Cooling method described in \cite{Ammann1997}. 

The ODT in our experiment 
has trapping frequencies of $\omega_{ODT,z} \simeq 2\pi \times 500$\,Hz and $\omega_{ODT,r} \simeq 2\pi \times 10$\,Hz which means that $\Delta t_{\text{col}}$ for the collimation pulse is only about $0.5\,$ms.  Fig.\ref{collimation}b shows our measurement of the axial width of the cloud after release as a function of TOF, with and without the collimation pulse. We observe that without the collimation pulse, the axial width exceeds 200\,$\mu$m within a few milliseconds, whereas with the ODT pulse, the axial width is only 70\,$\mu$m even after 25\,ms of TOF. Furthermore, one observes that the collimation is not perfect and the axial width continues to grow slowly. There are two main reasons responsible for this effect: First, due to its finite initial size the sample will always show dispersion. Second and more importantly, the magnetic potential which is focusing the gas in the radial direction is anti-confining in axial direction. This causes a small outward force on the particles. However, we still achieve our goal of limiting the axial width after long TOF. 

Due to the finite aspect ratio of our ODT, one has to consider the additional radial confinement created by the collimation pulse. Due to the short duration and the relatively weak strength of the radial confinement of the ODT, this effect is expected to be small and it only results in a change of the expansion time needed for focusing. 
When the collimation pulse is applied, the density of the sample does not decrease as quickly as when there is no collimation pulse (see Fig.\ref{collimation}). Therefore, scattering events during the expansion are more likely for the same interaction parameter. This issue can be addressed by choosing a magnetic field with even lower scattering length during the expansion with the collimation pulse.


\subsection{Magnification}
When the sample is condensed into low momentum states (Fig.\ref{fig:focusing}b), the optical density in the momentum space image is concentrated to a small central area. Access to the precise distribution in this region can be hindered by experimental limits to the optical resolution. There are several interesting phenomena, such as the phase coherence near the BKT phase transition, where it would be particularly desirable to resolve this region of interest. We therefore propose a method to magnify the mapped momentum distribution in order to improve momentum resolution without an improvement in optical resolution. 

\begin{figure} [ht!]
	\includegraphics [width= 8cm] {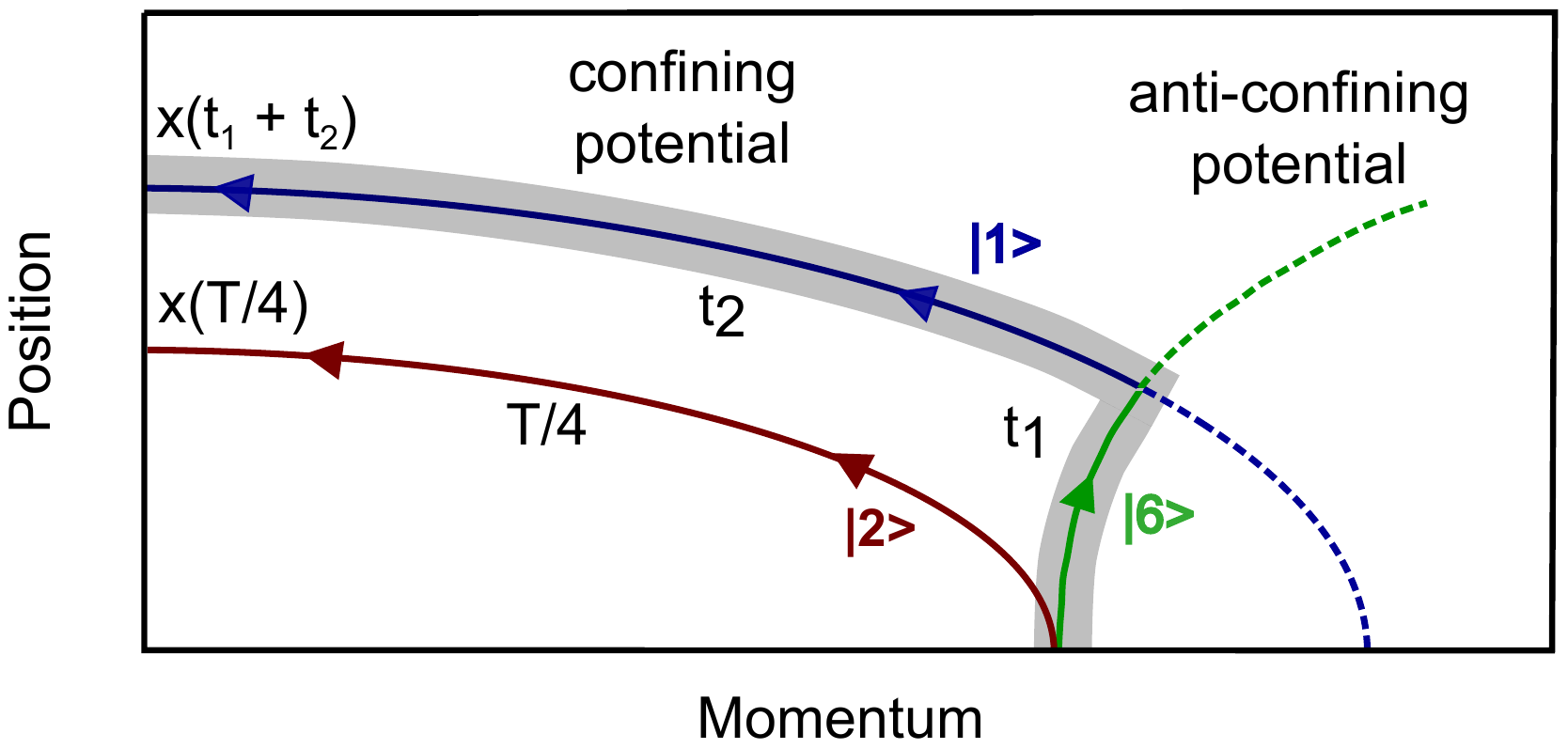}
\caption{Magnifying the momentum distribution. Phase-space trajectory of a particle traveling in a combination of anti-confining and confining potentials (\textit{grey shaded}). By letting the system expand first in a anti-confining potential, a given value of $p$ is mapped to a larger value of $x$, thus leading to an effective magnification. The colors represent the hyperfine levels of $^6$Li.} 
	\label{fig:magnification}
\end{figure}

In order to magnify the momentum distribution using only the focusing method, one could decrease $\omega_{\text{exp}}$ and then perform the focusing for the corresponding longer $T_{\text{exp}}/4$. However, the effectiveness of this method for magnification is limited because the magnification factor has a linear dependence on $\omega_{\text{exp}}$. 
Instead we propose to allow the gas to first expand in an anti-confining potential for a time $t_1$ and then let it evolve in a confining potential for a time $t_2$ (see Fig.\ref{fig:magnification}a).
In the language of optics, this is equivalent to placing a diverging lens in the optical path before a converging lens. By carefully choosing  $t_1$ and $t_2$, it is possible to obtain a magnified momentum distribution. This can be seen by solving the equations of motion for a classical particle traveling in the two potentials. We obtain the final position as a function of tunable parameters $t_1$ and $t_2$,

\begin{align}
x(t = t_1 + t_2) &= x_0 [\cos{\omega t_2}\cosh{\omega t_1} + \sin{\omega t_2}\sinh{\omega t_1}]\nonumber \\ 
& \quad+  \Big(\frac{p_0}{m\omega} \Big) [\exp{\omega t_1}].
\label{magnificationeq}
\end{align}\\
Here, $x_0$ and $p_0$ are the initial position and momentum of the particle. For simplicity we assume the two potentials to have the same curvature leading to the same $\omega$. We see that the effective magnification factor  is given by $M = \exp{\omega t_1}$ and thus grows exponentially with $t_1$. 
We further note from Eq.\ref{magnificationeq} that, the final position is generally dependent on both the initial momentum as well as the initial position. Therefore in order to obtain a pure mapping between momentum and position space, we find $t_2$ such that the first term in Eq.\ref{magnificationeq} vanishes,
\begin{equation}
\cos{\omega t_2}\cosh{\omega t_1} + \sin{\omega t_2}\sinh{\omega t_1} = 0.
\end{equation}
Thus, the final measured position is only a function of the initial momentum times the tunable magnification factor. 

To experimentally realize this technique, one can exploit the hyperfine structure of $^6$Li \cite{Gehm2003}. Initially the system consists of a mixture of atoms in the high-field seeking states $\vert 1 \rangle$ and $\vert 2 \rangle$. 
By applying a microwave $\pi$-pulse, one can transfer atoms in state $\vert 1 \rangle$ to a hyperfine state $\vert 6 \rangle = \vert F=3/2,m_F=3/2\rangle$ just before release. The atoms in state $\vert 6 \rangle$ are low-field seeking and thus experience a magnetic potential with opposite sign when released from the optical trap. After a time $t_1$ in this anti-confining potential, a second $\pi$-pulse can be applied to transfer the atoms in state $\vert 6 \rangle$ back to state $\vert 1 \rangle$. After allowing the gas to evolve in the now confining potential for a time $t_2$, one can then image the magnified 2D density distribution of the atoms in state $\vert 1 \rangle$. 

\section{Conclusions}

In this work we have established a set of methods for performing matter wave Fourier optics with strongly interacting quantum gases. Applying these techniques to a 2D quantum gas of $^6$Li atoms allowed us to directly observe the 2D momentum distribution of the system. 
Furthermore, we propose a technique to magnify the momentum distribution by letting the system evolve in a combination of anti-confining and confining potentials. In this way, large magnification factors can be achieved which would allow for detailed studies of the low-momentum region of a momentum distribution. 

Moreover, these techniques are not limited to bulk systems but can also be applied to other configurations such as optical lattices. By performing expansion for a time $T_{\text{exp}}/2$ instead of $T_{\text{exp}}/4$, it may be possible to access the topological phase of the sample, similar to 4f-imaging in optics. This will open up possibilities to perform matter wave phase-contrast imaging with quantum gases.

\section{Acknowledgments}
We thank Igor Boettcher for valuable discussions and support with the theoretical derivations. The authors gratefully acknowledge support from ERC starting Grant No. 279697, the Helmholtz Alliance HA216/EMMI, the Heidelberg Center for Quantum Dynamics, and the Landes\-graduierten\-f\"orderung  Baden-W\"urttemberg.

\appendix
\section{Time-evolution of quantum field operators in a harmonic potential}
 A quantum gas can be described in the second quantization formalism using quantum field operators $\hat{\Psi}(x)$ which set the occupation number of a particular quantum state in the position basis. For simplicity we only consider a 1D gas here. Due to the absence of interactions during the ballistic expansion, the expressions factorize and can thus easily be generalized to the 2D and 3D case. We show that for a gas evolving in a harmonic potential, the momentum distribution is mapped to the spatial density distribution after a quarter of the trap period. We construct bosonic field operators which obey the commutation relations
 \begin{equation}
 [\hat{\Psi}(x,t), \hat{\Psi}^\dagger (y,t)] = \delta(x - y).
 \label{commutation}
 \end{equation}
 The same can be done for fermionic operators but the obtained results remain unchanged. The time-evolution of the field operators is governed by the Heisenberg equation
 \begin{equation}
  i \hbar \,\partial_t \hat{\Psi}(x, t) = [\hat{\Psi}(x, t),\hat{H}(t)].
 \label{HeisenbergEq}
 \end{equation}
 Here, the Hamiltonian $\hat{H}(t)$ is constructed in second-quantization form according to
 \begin{eqnarray}
 \hat{H}(t) &=& H(\hat{\Psi}(x, t),\hat{\Psi}^{\dagger}(x, t)) \nonumber\\
  &=& \int dx \, \hat{\Psi}^{\dagger}\Big[-\frac{\hbar^2}{2m}\nabla^2 + \frac{1}{2}m\omega_{exp}^2x^2 \Big]\hat{\Psi}.
  \label{Hamiltonian}
 \end{eqnarray}
 Inserting this into Eq.\ref{HeisenbergEq} and using the commutation relation Eq.\ref{commutation}, we find that
 \begin{align}
 i \hbar \, \partial_t \hat{\Psi}(x, t) = \Big[-\frac{\hbar^2}{2m}\nabla^2 + \frac{1}{2}m\omega_{exp}^2x^2 \Big]\hat{\Psi}(x, t). 
 \label{Schrodinger}
 \end{align}
 This looks like the Schr\"{o}dinger equation but is the full evolution equation for the field operators. We expand $\hat{\Psi}(x)$ in terms of a time-dependent part and position-dependent Hermite Functions $H_n (\tilde{x})$ by using the ansatz
 \begin{equation}
 \hat{\Psi}(x,t) = \sum_n \hat{\psi}_n(t)\, H_n (\tilde{x}),
 \label{timedependence1}
 \end{equation}
 where $\tilde{x} = x/l_0$ with oscillator length $l_0 = \sqrt{\hbar/m\omega_{\text{exp}}}$. Using this ansatz in the field equation Eq.\ref{Schrodinger}, we obtain
 \begin{eqnarray}
 i\hbar\frac{\partial}{\partial t}\hat{\psi_n}(t)  &=& E_n \, \hat{\psi_n}(t), \nonumber  \\
 \hat{\psi_n} (t) &=& \hat{\psi_n}(0) \, e^{-iE_n t /\hbar}. 
 \end{eqnarray}
 Substituting this in Eq.\ref{timedependence1}, we get
 \begin{equation}
 \hat{\Psi}(x,t) = \sum_n \hat{\psi}_n(0)e^{-iE_n t /\hbar} H_n (\tilde{x}).
 \label{timedependence2}
 \end{equation}
 At $t = \frac{T_{\text{exp}}}{4} = \big( \frac{2\pi}{\omega} \big) \frac{1}{4}$ , where $T_{\text{exp}}$ is the time-period of the oscillator,
 \begin{equation*}
 E_n \cdot T_{\text{exp}}/4 = \hbar \omega_{\text{exp}} (n + \frac{1}{2}) \cdot\frac{\pi}{2\omega_{\text{exp}}} = \frac{\hbar\pi}{2}(n + \frac{1}{2}).
 \end{equation*}
 Since $e^{-i\pi/2} = -i$, Eq.\ref{timedependence2} simplifies to
 \begin{equation}
 \hat{\Psi}(x, T_{\text{exp}}/4) = \sum_n (-i)^n e^{-i\pi/4} H_n (\tilde{x}) \hat{\psi_n}(0).
 \end{equation}
 
 We now use the following rule for the Fourier transform of a Hermite function,
 \begin{equation}
 \tilde{H_n}(Y) = (-i)^n H_n(Y),
 \label{Hermiterule}
 \end{equation}
 to arrive at the field operator in momentum space
 \begin{eqnarray}
 \tilde{\Psi}(p, t) &=& \frac{1}{l_0}\int dx e^{ipx/\hbar} \hat{\Psi}(x,t) \nonumber \\
 &=& \sum_n \hat{\Psi}_n (t) (-i)^n H_n(\tilde{p})
 \label{Psimomentum}
 \end{eqnarray}
  with $\tilde{p} = pl_0/\hbar$. We then find
\begin{equation}
\hat{\Psi}^\dagger (x, T_{\text{exp}}/4) \hat{\Psi}(x, T_{\text{exp}}/4) \qquad  \qquad   \qquad  \qquad \nonumber 
\end{equation}
\begin{eqnarray}
&=&  \sum_{n,n'} (i)^n (-i)^{n'} H_n (x) H_{n'} (x) \hat{\psi_n}^\dagger(0)\hat{\psi_{n'}}(0) \nonumber \\
&=& \tilde{\Psi}^\dagger (p = m\omega x, 0) \tilde{\Psi}(p = m\omega x, 0),
\end{eqnarray}
 where we used $\tilde{p} = \tilde{x}$ in Eq.\ref{Psimomentum}. Therefore, after $T_{\text{exp}}/4$, we obtain the Fourier transform of the initial field operator. The corresponding spatial density  is given by
 \begin{align}
 n(x, T_{\text{exp}}/4) &= \langle \hat{\Psi}^\dagger (x, T_{\text{exp}}/4)\hat{\Psi}(x, T_{\text{exp}}/4) \rangle \nonumber\\
 &= \langle  \tilde{\Psi}^\dagger (p,0)\tilde{\Psi}(p,0) \rangle = n(p,0),
 \end{align}
 with $p = m\omega x$. This proves that the spatial density distribution after $t_{exp} = T_{\text{exp}}/4$ reflects precisely the initial momentum distribution with a scaling factor $1/m\omega_{\text{exp}}$.
 
\section{Collisions in an expanding 2D gas}
\begin{figure} [htb]
	\includegraphics [width= 7cm] {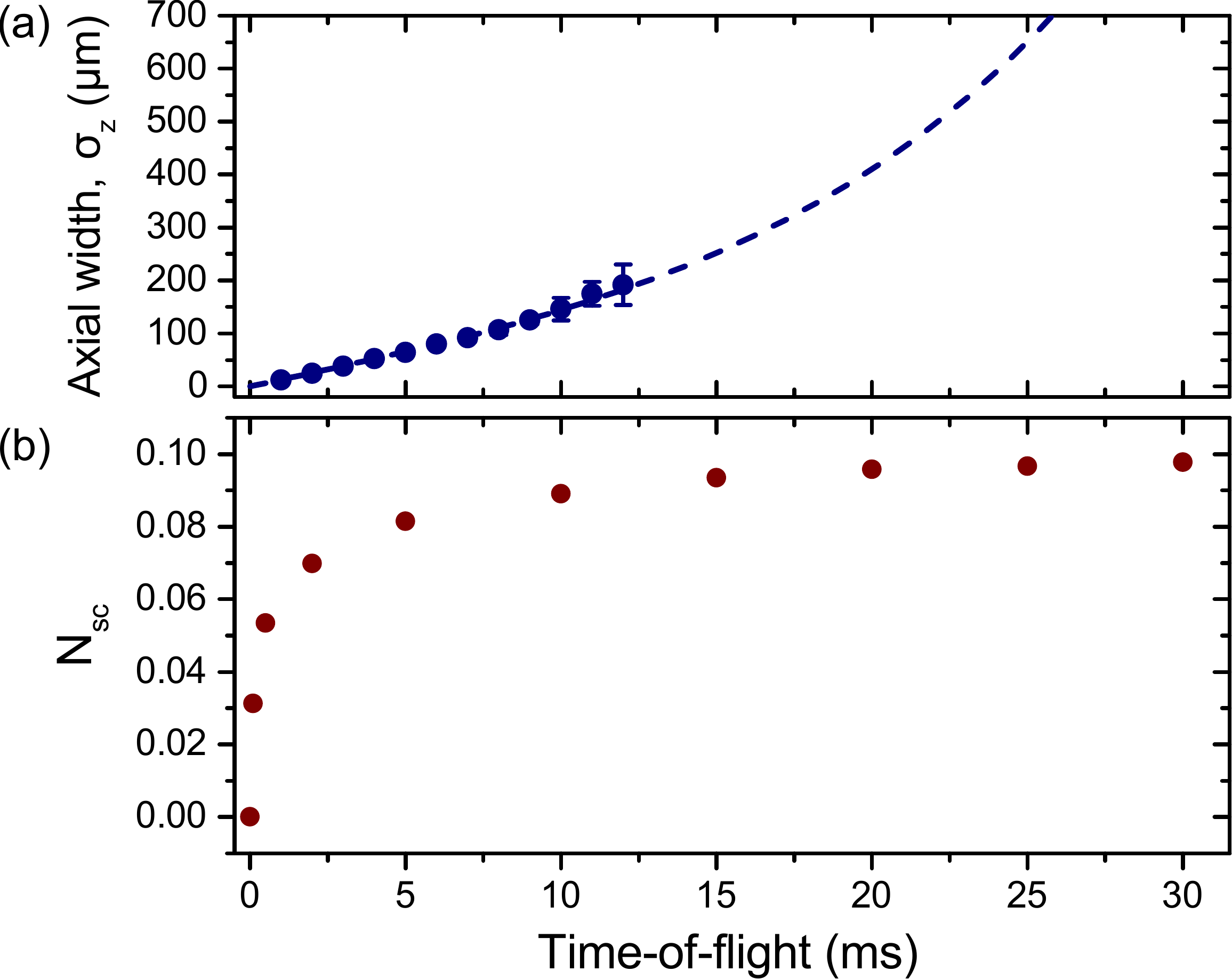}
\caption{(a) Experimentally obtained axial width and the corresponding theoretical prediction for a cloud expanding in the magnetic potential at $692\,$G. 
(b) Average number of scattering events per particle in the sample as a function of the elapsed expansion time. After a quick initial rise of $N_{\text{sc}}$,  the number of scattering events saturates and reaches about $0.1$ for expansion times exceeding $20\,$ms.}
\label{fig:scattering}
\end{figure}

We estimate the number of scattering events that occur while the particles are expanding using the expression
\begin{equation}
\Gamma = n \sigma v,
\label{scatteringrate}
\end{equation} 
where n is the density, $\sigma$ is the scattering cross-section and $v$ is the mean velocity of the particles. In our case, we want to investigate the scattering events for a sample with a temperature of about $60\,$nK at a magnetic offset field of $692\,$G, where the images in Fig.\ref{fig:focusing} are taken. At this field, the atoms are paired up and form deeply bound bosonic molecules. Their cross-section $\sigma$ is thus given by $\sigma= 8 \pi a^2_{\text{mol}}$, where $a_{\text{mol}}$ is the 3D scattering length between molecules which is $a_{\text{mol}} (692\,$G$)= 0.6 \times 1463 \,a_{\text{Bohr}}$ \cite{Petrov2005,Zuern2013}. One thus obtains $\sigma\simeq5.4\times10^{-14}\,$m$ ^2$. We assume a Maxwell-Boltzmann distribution of the velocities and obtain the following mean velocity $v=\sqrt{2 k_B T/m_{\text{mol}}}$, where $k_B$ is the Boltzmann constant, $T$ is the temperature of the sample and $m_{\text{mol}}= 2\, m_{\text{Li}^6}$ is the mass of a bosonic dimer. Hence, for our temperature of about $60\,$nK we obtain $v\simeq 0.0091\,$m/s.

To estimate the average scattering rate per particle in our trap, we evaluate the measured in-situ density distributions to calculate the average density 
and obtain $\bar{n}_{3D}\simeq 1.2 \times 10^{12}\,$cm$^{-3}$.
Using these numbers, we obtain an average in-situ scattering rate of $\Gamma_0 = \bar{n}_{3D}  \sigma v \simeq 580\,$Hz per particle. 

%

To calculate the number of scattering events during the expansion, we simulate the evolution of the gas in the weak magnetic potential using the method given in \cite{Hu2004,Ketterle2008}. This yields the axial width of the cloud $\sigma_z (t)$ as a function of the expansion time. The resulting widths together with the corresponding experimental data are shown in Fig.\ref{fig:scattering}a. Although this method was originally devised to describe the expansion of an anisotropic condensate in the 3D BEC-BCS crossover, the good agreement between the data and the prediction justifies its application to our system \footnote{Since we are in the BEC regime ($1/k_F a_{3D}>1$), the effective exponent $\gamma$ in the model described in \cite{Hu2004,Ketterle2008} can be set to $1$.}. Neglecting the motion in the radial direction, the density of the cloud is inversely proportional to the axial cloud width $\sigma_z$ and hence the scattering rate at an expansion time $t$ is given by
\begin{equation}
\Gamma(t) = \Gamma_0 \cdot \frac{\sigma_z (0)}{\sigma_z (t)}.
\label{sigma_t}
\end{equation}
By integrating this quantity over the complete expansion time of $25\,$ms, we finally obtain the average number of scattering events per particle in our sample during the expansion
\begin{equation}
N_{\text{sc}} (t=25\,\text{ms}) = \int_0 ^{t=25\,\text{ms}} \Gamma(t) dt \simeq 0.09.
\label{integral}
\end{equation} 
Fig.\ref{fig:scattering}a shows this quantity as a function of the expansion time. One observes that for $t=25\,$ms less than 10\,$\%$ of all particles undergo collisions during the focusing. It is however interesting to note that while half of the scattering events occur during the first $\simeq 0.5\,$ms there are still residual scattering events up to the complete expansion time of $25\,$ms.


\bibliography{References_momentum_imaging2}

\end{document}